\documentclass[10pt]{article}

\usepackage{amssymb}
\def\lddots{\mathinner{\mkern1mu\raise1pt\hbox{.}\mkern2mu
\raise4pt\hbox{.}\mkern2mu\raise7pt\vbox{\kern7pt\hbox{.}}\mkern1mu}} 
\makeatletter
\def\numberbysection{\@addtoreset{equation}{section}
\def\theequation{\thesection.\arabic{equation}}}
\makeatother

\numberbysection

\newcommand{\be}{\begin{eqnarray}}
\newcommand{\ee}{\end{eqnarray}}
\newcommand{\non}{\nonumber}

\begin{document}

\begin{titlepage}
\vskip 0.4cm
\strut\hfill
\vskip 0.8cm
\begin{center}


{\LARGE Boundary non-local charges from the open spin chain}

\vspace{10mm}

{\Large Anastasia Doikou\footnote{e-mail: doikou@lapp.in2p3.fr}}

\vspace{14mm}

\emph{ Laboratoire d'Annecy-le-Vieux de Physique Th{\'e}orique\\
LAPTH, CNRS, UMR 5108, Universit{\'e} de Savoie\\
B.P. 110, F-74941 Annecy-le-Vieux Cedex, France}

\end{center}

\vfill

\begin{abstract}

The $N$ site open XXZ quantum spin chain with a right non-diagonal boundary and special diagonal left boundary is considered. The boundary non-local charges originally obtained from a field theoretical viewpoint, for the sine Gordon model on the half line, are recovered from the spin chain point of view. Furthermore, the symmetry of the open spin chain is exhibited. More specifically, we show that certain non--local charges commute with the transfer matrix of the open spin chain, depending on the choice of boundary conditions. In addition, we show explicitly that for a special choice of the left boundary, one of the non--local charges, in a particular representation, commutes with each one of the generators of the blob algebra, and hence with the corresponding local Hamiltonian.

\end{abstract}

\vfill
\rightline{LAPTH-1029/04}
\baselineskip=16pt
\end{titlepage}


\section{Introduction}

Quantum groups have attracted much interest recently because of their intrinsic mathematical interest 
\cite{jimbo}--\cite{cha}, but also because they provide a powerful scheme for deriving solutions
of the Yang-Baxter equation \cite{baxter}--\cite{korepin} 
\be
R_{12}(\lambda_{1}-\lambda_{2})\ R_{13}(\lambda_{1})\
R_{23}(\lambda_{2}) =R_{23}(\lambda_{2})\ R_{13}(\lambda_{1})\ R_{12}(\lambda_{1}-\lambda_{2}). \label{YBE} \ee  $R$ acts on ${\mathbb V}^{\otimes 2}$, whereas the Yang--Baxter equation on ${\mathbb V}^{\otimes 3}$, and as customary $~R_{12} = R \otimes {\mathbb I}$, $~R_{23} ={\mathbb I} \otimes R$. The $R$ matrix may be attained by means of the quantum group
approach \cite{kure, jimbo2} by solving certain linear intertwining relations involving the $R$ matrix and coproducts of the quantum group generators.

Recently it was realized that a similar procedure may be
implemented for solving the reflection equation \cite{cherednik}, \begin{equation}
R_{12}(\lambda_{1}-\lambda_{2})\ {\cal K}_{1}(\lambda_{1})\
R_{21}(\lambda_{1}+\lambda_{2})\
{\cal K}_{2}(\lambda_{2})=
{\cal K}_{2}(\lambda_{2})\ R_{12}(\lambda_{1}+\lambda_{2})\
{\cal K}_{1}(\lambda_{1})\ R_{21}(\lambda_{1}-\lambda_{2}) \label{re} \end{equation} where ${\cal K}$ acts on ${\mathbb V}$, $~{\cal K}_{1} = {\cal K} \otimes {\mathbb I}$, $~{\cal K}_{2} ={\mathbb I} \otimes {\cal K}$, and $R$ is a solution of the Yang--Baxter equation (\ref{YBE}). The Yang--Baxter and reflection equations offer collections of algebraic constraints obeyed by both boundary integrable field theories and lattice integrable models. The first time that boundary non-local charges were constructed was in the context of the boundary sine-Gordon model in the `free fermion' point \cite{mene}, while representations of such objects were found in \cite{neg} for higher rank algebras. Later non-local charges were derived for the affine Toda field theories with particular boundary conditions \cite{dema}. Note that as in the case of the Yang--Baxter equation, the derivation of the ${\cal K}$ matrix also requires the solution of linear intertwining relations that involve elements of the reflection algebras, defined by (\ref{re}), and ${\cal K}$ \cite{dema}--\cite{bako}.

In this article we deal with the generalized $N$ site XXZ open spin chain with a non-diagonal right boundary and a special diagonal left boundary. In particular, we consider the case in which each site of the spin chain is associated to a copy of 
$U_{q}(\widehat{sl_{2}})$, i.e. we deal with a purely algebraic construction. It will be instructive to write down the Hamiltonian of the spin 1/2 XXZ model with the aforementioned boundary conditions, i.e. \be {\cal H}= -{1\over 4}
\sum_{i=1}^{N-1}\Big (\sigma_{i}^{x}\sigma_{i+1}^{x} +\sigma_{i}^{y}\sigma_{i+1}^{y}+ \cosh
i\mu\ \sigma_{i}^{z}\sigma_{i+1}^{z}\Big )- {1\over 4}\sinh i\mu\ (\sigma_{N}^{z}-\sigma_{1}^{z})-{N+1\over 4}\ \cosh i\mu \non\\ -{\sinh i\mu \ \sinh(im \mu)\over 4 \sinh i \mu({m\over 2}+\zeta)\ \cosh i \mu({m\over 2}-\zeta)}\ \sigma_{1}^{z} +{\sinh i\mu \over 4 \sinh i \mu({m\over 2}+\zeta)\ \cosh i \mu ({m\over 2}-\zeta)}\
\sigma_{1}^{x} +c_{1} +c_{2}\ \sigma_{N}^{z}.
\label{Hbound} \ee The constants $m,\ \zeta$ are dictated by integrability as will be clear later, and $c_{1}, \ c_{2}$ depend on the choice of the left diagonal boundary. Notice that the values $\zeta = -{m\over 2}$, $~\zeta = {m\over 2}+{\pi \over 2 \mu}\  ~(mod({\pi \over \mu}))$ merit special treatment. 
In fact, for these particular values the contribution of the right boundary is proportional to one, and therefore the first two terms of the second line of the Hamiltonian (\ref{Hbound}) (proportional to $\sigma_{1}^{z}$, $~\sigma_{1}^{x}$) are replaced by a constant.

As is well known \cite{sklyanin} the open spin chain may be constructed by using a generalized solution of the reflection equation denoted as ${\cal T}$. The asymptotic behaviour of ${\cal T}$ as $\lambda \to \infty$ gives rise to boundary non--local charges, which turn out to be coproducts of generators of the reflection algebra \cite{mene, dema}. Linear intertwining relations involving such generators bear algebraic Bethe ansatz type exchange relations, enabling the study of the corresponding transfer matrix symmetry. Using these relations we are able to show that certain combinations of the aforementioned non--local charges commute with the  transfer matrix of the open spin chain depending on the choice of the left boundary. This is the first time to our knowledge that such exchange relations have been presented, and such method has been employed for the investigation of the symmetry of an open integrable spin chain with non-diagonal boundaries. We should  emphasize that these findings rely on algebraic considerations, and therefore they are independent of the choice of representation. Finally, one more new key result is presented, that is the commutation of a particular representation of one of the  boundary charges with each one of the generators of the blob algebra \cite{tl}--\cite{doma} in the XXZ representation. Using this piece of information we show that the corresponding local Hamiltonian commutes with the boundary non-local charge as well. The open XXZ spin chain is used as a paradigm, nevertheless the results of this investigation may be generalized for models associated to higher rank algebras.
      
The present approach rests primarily on the fact that we have at our disposal solutions of the reflection equation by other means (e.g Hecke algebraic approach) and we use them to extract the boundary non--local charges by studying the asymptotics of the general solution ${\cal T}$. A generalization for the $U_{q}(\widehat{sl_{n}})$ case with boundary conditions different to the ones appearing in \cite{dema}, is also a main motivation for this investigation. The boundary non-local charges have not been determined yet in this case from the field theory point of view, because such boundary conditions have not been studied classically, so the starting point should better be the affine Hecke algebra \cite{cher, male}. 

\section{Constructing spin chains}

An integrable theory possesses a set of mutually commutative operators. In the presence of boundaries, the integrability is not ensured, as the commutativity may not apply anymore. However, there are cases where one may find integrable boundary conditions, and it is possible to modify the set of commutative operators \cite{sklyanin}. In particular, in the generalized QISM for open spin chains, introduced in \cite{sklyanin}, two building blocks are employed, namely the $R$ (${\cal L}$) and ${\cal K}$ matrices. Once having these constructing elements one may derive the transfer matrix of the open spin chain, which gives rise to the set of mutually commuting operators.
Let us now briefly review some basic definitions regarding the $R$ (${\cal L}$) and ${\cal K}$ matrices for the well known XXZ model.

\subsection{$R$ matrix and Lax operator}

Let us first recall the XXZ spin $1/2$ $R$ matrix, which is a solution of the Yang-Baxter equation (\ref{YBE}). It is simply the $R$ matrix associated to the fundamental representation of $U_{q} (sl_{2})$, acting on $({\mathbb C}^{2})^{\otimes 2}$. It is given below in the so called homogeneous gradation
\be R(\lambda) = \left(
\begin{array}{cc}
\sinh(\lambda +{i\mu\over 2} +{i\mu \sigma^z \over 2})   &\sigma^{-} e^{\lambda}\sinh i \mu  \\
\sigma^{+}  e^{-\lambda}\sinh i \mu   &\sinh(\lambda +{i\mu  \over 2} -{i\mu \sigma^z \over 2})     
\end{array} \right)\label{r} \ee  
where $\sigma^{z},\ \sigma^{\pm}$ are the usual $2 \times 2$ Pauli matrices. The $R$ matrix in the principal gradation can be deduced by means of a simple gauge transformation, i.e.
\be R_{12}^{(p)}(\lambda) = {\cal V}_{1}(\lambda)\ R_{12}^{(h)}(\lambda)\ {\cal V}_{1}(-\lambda), ~~~~~{\cal V}(\lambda) = diag(1,\ e^{\lambda}). \label{gauge1} \ee  Having specified $R \in \mbox{End}({\mathbb C}^2 \otimes {\mathbb C}^2)$ one may derive a more general object ${\cal L} \in \mbox{End}({\mathbb C}^2) \otimes {\cal A}$, where the algebra ${\cal A}=U_{q}(\widehat{sl_{2}})$ is defined by the fundamental algebraic relation \be R_{12}(\lambda_{1} -\lambda_{2})\ {\cal L}_{13}(\lambda_{1})\ {\cal L}_{23}(\lambda_{2})= {\cal L}_{23}(\lambda_{2})\ {\cal L}_{13}(\lambda_{1})\ R_{12}(\lambda_{1} -\lambda_{2}). \label{funda}\ee 
A simple solution of the latter equation, which we shall use hereafter, may take the following form written in the homogeneous gradation
\be {\cal L}(\lambda) =    \left(
\begin{array}{cc} 
\sinh(\lambda +{i\mu\over 2} +i\mu\ h_{1} )   &  e^{\lambda}\ f_{1}\ \sinh i \mu  \\
e^{-\lambda}\ e_{1}\ \sinh i \mu   & \sinh(\lambda +{i\mu  \over 2} - i\mu\ h_{1} )   \end{array} \right),  \label{l} \ee where $k_{1}= q^{h_{1}}$, $e_{1}$, $f_{1}$ are the generators of $U_{q}(sl_{2})$. A more detailed description of $U_{q}(\widehat{sl_{2}})$ will be presented in the Appendix. ${\cal A}$ is endowed with a coproduct, $\Delta: {\cal A} \to {\cal A} \otimes {\cal A}$, in fact the entries of ${\cal L}$, which generate ${\cal A}$, form the coproducts \be \Delta({\cal L}_{ij}(\lambda)) = \sum_{k=1}^{2}{\cal L}_{kj}(\lambda) \otimes {\cal L}_{ik}(\lambda), ~~~~i,\ j \in \{1,\ 2\}. \label{cob1} \ee It will be also useful for the following to define $\Delta': {\cal A} \to {\cal A}$ which may be attained from $\Delta$ by permutation. Indeed, let $\Pi: u \otimes v \to v \otimes u$ then \be \Delta'= \Pi \circ \Delta \label{cc} .\ee
The $L$ coproduct may be also derived by recursion as \be \Delta^{(L)} = (\mbox{id} \otimes \Delta^{(L-1)})\Delta, ~~~~\Delta^{'(L)} = (\mbox{id} \otimes \Delta^{(L-1)})\Delta'. \label{co2a} \ee 
In general by considering tensor products of $N$ ${\cal L}$ matrices we may derive the algebraic monodromy matrix $T \in \mbox{End}({\mathbb C}^{\otimes 2}) \otimes {\cal A}^{\otimes N}$ \be T_{0}(\lambda) = (\mbox{id} \otimes \Delta^{(N)}){\cal L}(\lambda) = {\cal L}_{0N}(\lambda)\ {\cal L}_{0\ N-1}(\lambda) \ldots {\cal L}_{01}(\lambda) \label{tt} \ee where now the subscripts $a \in \{1,\ldots, N\}$ refer to the site of the coproduct sequence. Usually, the indexes $a$ associated to the `quantum spaces' are suppressed from the monodromy matrix, whereas the index 0, which corresponds to the `auxiliary space' is kept. To distinguish the indexes appearing in (\ref{cob1}) and (\ref{tt}) we shall hereafter denote $a,\ b$ the indexes that refer to the site of a coproduct sequence as in (\ref{tt}),  and $i,\ j$ to the ones referring to the entries of the matrix as in (\ref{cob1}).
Finally, by taking the trace over the auxiliary space we define the transfer matrix $t \in {\cal A}^{\otimes N}$ \be t(\lambda) =Tr_{0}\ \{ T_{0}(\lambda) \}, \label{transfer1} \ee which provides a family of commuting operators \be \Big [ t(\lambda),\ t(\lambda')\Big ]=0, \ee ensuring the integrability of the model. Note that the `spin chain' described by  (\ref{tt}), (\ref{transfer1}) is an algebraic construction consisting of $N$ copies of ${\cal A}$, since the quantum spaces are not represented but they are copies of ${\cal A}$. The aforementioned algebraic construction may be interpreted as a physical spin chain once each one of the $N$ copies of ${\cal A}$ is mapped to some finite or infinite dimensional space. Then the spectrum of the transfer matrix may be derived, and all the physically relevant quantities can be computed.

\subsection{The ${\cal K}$ matrix}

Our aim now is to construct the open spin chain. For this purpose we need to consider one more fundamental object, namely the ${\cal K}$ matrix, which  is a solution of the reflection equation (\ref{re}). As proposed in \cite{male, doma} an effective way of finding solutions of the reflection equations (\ref{re}) is by exploiting certain
algebraic structures such as Temperley--Lieb and blob algebras \cite{tl}--\cite{doma}.

The blob algebra $b_N(q,Q)$, which is a quotient of the affine Hecke algebra \cite{cher}, is defined by generators ${\cal U}_{1},{\cal U}_{2},...,{\cal U}_{N-1}$ and ${\cal U}_{0}$, and
relations:
\begin{eqnarray} {\cal U}_{l}\ {\cal U}_{l} &=& \delta\ {\cal U}_{l},~~~~{\cal U}_{0}\
{\cal U}_{0} = \delta_{0}\ {\cal U}_{0} \non\\
{\cal U}_{l\pm 1}\ {\cal U}_{l}\ {\cal U}_{l\pm 1} &=& {\cal U}_{l\pm 1}, ~~~~{\cal
U}_{1}\ {\cal U}_{0}\ {\cal U}_{1} = \gamma\ {\cal U}_{1} \non\\
\Big [ {\cal U}_{l},\ {\cal U}_{k} \Big ] &=& 0, ~~~~|l-k| \neq 1
\label{TL}
\end{eqnarray} $\delta=-(q+q^{-1})$, $q=e^{i\mu}$, and $\delta_{0}$, $\gamma$ are constants depending on $q$ and $Q=e^{im \mu}$.\\
We give here the generators ${\cal U}_{l},\ l\in \{1,..,N-1 \}$, ${\cal U}_{0}$ of the blob algebra in the XXZ representation, i.e. let the tensor representation $h: b_{N}(q,Q) \to \mbox{End}((\mathbb C^{2})^{\otimes N})$ such that (see also \cite{masa})
\be h({\cal U}_{l}) &=& 1 \otimes \ldots \otimes \left(
\begin{array}{cccc}
    0    &0        &0       &0   \\
    0    &-q       & 1      &0   \\
    0    &1        &-q^{-1} &0   \\
    0    &0        &0       &0
\end{array} \right) \otimes \ldots \otimes 1,\non\\ h({\cal
U}_{0}) &=&\left( \begin{array}{cc}
      -Q      & 1        \\
       1       &-Q^{-1}    \\
\end{array} \right)
\otimes \ldots \otimes 1 \label{tlg} \ee with $h({\cal U}_{l})$ acting non-trivially on ${\mathbb V}_{l} \otimes {\mathbb V}_{l+1}$, (where ${\mathbb V} =\mathbb
C^{2}$) and $h({\cal U}_{0})$ acting on ${\mathbb V}_{1}$. Note that $h({\cal U}_{l})$ $l\in \{1, \ldots, N-1 \}$ are actually the XXZ representations of the Temperley-Lieb algebra generators \cite{tl, mawo, doma}. 

As argued in \cite{doma} tensor representations of the blob algebra provide solutions of the reflection equation. Hence a solution of the reflection equation (\ref{re}) may be written in terms of the blob algebra generator $h({\cal U}_{0})$ (\ref{tlg}) as\footnote{The notation here is slightly modified compared to \cite{doma}. Also, for this representation of the blob algebra, we find: $\delta_{0} = -(Q+Q^{-1})$ and $\gamma = qQ +q^{-1}Q^{-1}$.}  \be
{\cal K}^{(b)}(\lambda) = 2\sinh (\lambda-{i m \mu\over 2}-i\mu\zeta)\ \cosh
(\lambda-{i m\mu \over 2}+i\mu\zeta)\ {\mathbb I} + \sinh 2\lambda\ h({\cal U}_{0}). \label{ansatz1} \ee Recall also that the XXZ spin $1/2$ $R$ matrix (\ref{r}) may be also written in terms of the representation $\rho({\cal U}_{l})$, \cite{jimbo}, i.e. \be R_{12}(\lambda) = {\cal P}_{12}(\sinh (\lambda +i\mu)\ {\mathbb I} +\sinh \lambda\ h({\cal U}_{1})) \label{ansatz2} \ee where ${\cal P}$ is the permutation operator such that ${\cal P}(a\otimes b)= b \otimes a$. 

Let us now explain why the matrix obtained from the blob algebra (\ref{ansatz1}) and the ${\cal K}$ matrix \cite{DVGR, GZ} (written in the homogeneous gradation) coincide, (written in the homogeneous gradation) coincide, subject to certain identifications. The ${\cal K}$ matrix found in \cite{DVGR, GZ} may be written as \be && {\cal K}_{11}(\lambda) =e^{2\lambda}{e^{i\mu \xi} \over 2\kappa} -{e^{-i\mu \xi} \over 2\kappa},~~{\cal K}_{22}(\lambda)=e^{-2\lambda}{e^{i\mu \xi} \over 2\kappa} -{e^{-i\mu \xi} \over 2\kappa} \non\\
&&{\cal K}_{12}(\lambda) ={\cal K}_{21}(\lambda) = \sinh 2  \lambda. \label{k} \ee  The ${\cal K}$ matrix from the blob algebra (\ref{ansatz1}) on the other hand can be written in an explicit form by using (\ref{tlg}):
\be&& {\cal K}^{(b)}_{11}(\lambda) =-e^{2 \lambda} \sinh i \mu m -\sinh 2 i \mu \zeta, ~~{\cal K}^{(b)}_{22}(\lambda)=-e^{-2\lambda}\sinh i \mu m- \sinh 2i \mu \zeta \non\\
&&{\cal K}^{(b)}_{12}(\lambda) ={\cal K}^{(b)}_{21}(\lambda) = \sinh 2  \lambda.  \label{22} \ee It is now clear that the above expressions for ${\cal K}$ (\ref{k}) and ${\cal K}^{(b)}$ (\ref{22}) coincide subject to the following identifications 
\be {e^{i\mu \xi} \over 2\kappa} = -\sinh i m \mu, ~~~{e^{-i\mu \xi} \over 2\kappa}= \sinh 2i \mu \zeta. \label{ide}\ee 
The blob algebra (affine Hecke algebra \cite{cher} for e.g. $U_{q}(\widehat{sl_{n}})$) as already mentioned is considered as the starting point in the present analysis providing $c$-number solutions (\ref{ansatz1}) of (\ref{re}). Once having solutions of the reflection equation, attained algebraically, one may build the corresponding open spin chain and derive certain boundary non-local charges \cite{mene, dema} as will see in the subsequent sections. Note finally that the ${\cal K}$ matrix in the principle gradation can be obtained by the gauge transformation \be {\cal K}^{(p)}(\lambda)= {\cal V}(\lambda)\ {\cal K}(\lambda)\ {\cal V}(\lambda). \label{gauge2} \ee  

\subsection{The underlying algebra and the open spin chain}

Having specified c-number solutions of the (\ref{re}) we may build the more general solution of (\ref{re}) as argued in \cite{sklyanin}. To do so it is necessary to define the following object \be \hat {\cal L}(\lambda) = {\cal L}^{-1}(-\lambda), \ee in particular
\be \hat {\cal L}(\lambda) =    \left(
\begin{array}{cc}
\sinh(\lambda +{i\mu\over 2} +i\mu\ h_{1} )   &  e^{-\lambda}\ f_{1}\ \sinh i \mu  \\
e^{\lambda}\ e_{1}\ \sinh i \mu   & \sinh(\lambda +{i\mu  \over 2} - i\mu\ h_{1} )  
\end{array} \right),  \label{l´} \ee The general solution of (\ref{re}) is then  given by \cite{sklyanin}:
\be {\mathbb K}(\lambda) = {\cal L}(\lambda-\Theta)\  ({\cal K}(\lambda)\otimes {\mathbb I})\  \hat {\cal L}(\lambda +\Theta), \label{gensol} \ee where ${\cal K}$ is the c-number solution of the reflection equation, $\Theta$ is called inhomogeneity and hereafter for simplicity we shall consider it to be zero. ${\mathbb K}$ generates the elements of the reflection algebra ${\mathbb R}$, which obey exchange relations dictated by the algebraic constraints (\ref{re}) (see also \cite{sklyanin}). It is clear that the general solution (\ref{gensol}) allows the  asymptotic expansion as $\lambda \to \infty$ as we shall see in subsequent sections. The first order of such expansion ($\lambda$ independent) yields the generators of the boundary quantum group ${\mathbb B}(U_{q}(sl_{2}))$ (associated to the $U_{q}(sl_{2}))$ (\ref{l})), which obey commutation relations dictated by the defining relations (\ref{re}) as $\lambda_{i} \to \infty$. The boundary quantum group is a subalgebra of $U_{q}(sl_{2})$ and ${\mathbb R}$, and provides the underlying algebraic structure in the reflection equation exactly as quantum groups do in the Yang--Baxter equation.

The reflection algebra is also endowed with a coproduct inherited from ${\cal A}$. In particular, let us first derive the coproduct of $\hat {\cal L}$, i.e. \be\Delta(\hat {\cal L}_{ij}(\lambda)) = \sum_{k=1}^2 \hat {\cal L}_{ik}(\lambda) \otimes \hat  {\cal L}_{kj}(\lambda)~~~~i,~j\in\{1, \ 2\}. \label{cob2} \ee Then it is clear from (\ref{cob1}), (\ref{cob2}) that the elements of ${\mathbb R}$ form coproducts $\Delta: {\mathbb R} \to {\mathbb R} \otimes {\cal A}$, such that (see also \cite{dema}) \be \Delta({\mathbb K}_{ij}(\lambda)) = \sum_{k,l=1}^2{\mathbb K}_{kl}(\lambda) \otimes {\cal L}_{ik}(\lambda) \hat {\cal L}_{lj}(\lambda)~~~~i,~j \in \{1, \ 2\}. \label{coc} \ee Our final aim is to build  the corresponding quantum system that is the open quantum spin chain. To achieve that we shall need tensor product realizations of the general solution (\ref{gensol}). We first define \be \hat T_{0}(\lambda) = (\mbox{id} \otimes \Delta^{(N)}) \hat {\cal L}(\lambda) =  \hat {\cal L}_{01}(\lambda)\ldots  \hat {\cal L}_{0N}(\lambda) \label{th} \ee then the  general solution of the (\ref{re}) takes the form \be {\cal T}_{0}(\lambda) = T_{0}(\lambda)\ {\cal K}_{0}^{(r)}(\lambda)\ \hat T_{0}(\lambda),  \label{transfer0} \ee where the entries of ${\cal T}$\footnote{The operator ${\cal T}$ in principal and homogeneous gradation are related via the gauge transformation (\ref{gauge1}) \be {\cal T}_{0}^{(p)}(\lambda)={\cal V}_{0}(\lambda)\ {\cal T}_{0}(\lambda)\ {\cal V}_{0}(\lambda) \label{tp} \ee} are simply coproducts of the elements of the reflection algebra, namely \be {\cal T}_{ij}(\lambda)= \Delta^{(N)}({\mathbb K}_{ij}(\lambda)). \label{tc} \ee  Finally we introduce the transfer matrix of the open spin chain \cite{sklyanin}, which may be written as \be t(\lambda) = Tr_{0}\ \Big \{ M_{0}\ {\cal K}_{0}^{(l)}(\lambda)\ {\cal T}_{0}(\lambda)\Big \}.  \label{transfer} \ee ${\cal K}^{(r)}$ is a solution of the reflection equation (\ref{re}) and $~{\cal K}^{(l)}(\lambda) = {\cal K}(-\lambda -i \mu)^{t}$ with ${\cal K}$ being also a solution of (\ref{re}), not necessarily of the same type as ${\cal K}^{(r)}$. Also, we define \be && M={\mathbb I}, ~~~\mbox{principal gradation},\non\\ && M=diag(e^{i\mu},\ e^{-i\mu}), ~~~ \mbox{homogeneous gradation}. \ee
It can be shown \cite{sklyanin} using the fact that ${\cal T}$ satisfies the reflection equation, that the transfer matrix (\ref{transfer}) provides a family of commuting operators i.e., \be \Big [t(\lambda),\ t(\lambda')\Big ] =0, \label{com} \ee a fact that guarantees the integrability of the model.

\section{Boundary non-local charges}

In this section the boundary non-local charges are obtained via the standard procedure, by considering the asymptotics of the operator ${\cal T}$ as $\lambda \to \infty$ (see also \cite{dema}). Henceforth, we shall consider ${\cal K}^{(l)} = {\mathbb I}$, ${\cal K}^{(r)} = {\cal K}$ given by (\ref{k}). If we consider a different choice for the left boundary we shall explicitly state it. Also, in the following  ${\cal L}$, $\hat {\cal L}$ and $T$( $\hat T$), ${\cal T}$ are treated as $n \times n$ matrices with entries being elements of ${\cal A}$, ${\cal A}^{\otimes N}$ respectively.\\
\\
$\bullet$ {\it Homogeneous gradation}: Let us start our analysis considering the asymptotics of  ${\cal L}$, $\hat {\cal L}$ given by ({\ref{l}) (\ref{l´}) respectively, namely (recall $q= e^{i\mu}$, and for the asymptotic study $\mu$ is taken to be finite. Also for simplicity the `auxiliary' space index $0$ is suppressed from $T$, $\hat T$ and later from ${\cal T}$ (\ref{transfer})) \be {\cal L}(\lambda \to \infty) \propto  \left(
\begin{array}{cc}
k_{1} &w\ f_{1} \\
   &  k_{1}^{-1} 
\end{array} \right), ~~~\hat {\cal L}(\lambda \to \infty) \propto   \left(
\begin{array}{cc}
k_{1} & \\
 w\ e_{1}  &  k_{1}^{-1} 
\end{array} \right) \label{l1} \ee with $~w=2q^{-{1\over 2}} \sinh i \mu$. Now it is straightforward to see by virtue of the form of ${\cal L}^+$ and $\hat {\cal L}^+$ that the monodromy matrices also reduce to upper, lower triangular matrices as $\lambda \to \infty$, \be T(\lambda \to \infty) \propto  \left(
\begin{array}{cc}
K_{1}^{(N)} & w\ F_{1} \\
          & (K_{1}^{(N)})^{-1}
\end{array} \right), ~~~\hat T(\lambda \to \infty) \propto  
\left(
\begin{array}{cc}
 K_{1}^{(N)}& \\
 w\ E_{1}^{(N)}  &  (K_{1}^{(N)})^{-1}
\end{array} \right) \label{T1} \ee 
with $K_{1}^{(N)}$, $E_{1}^{(N)}$, $F_{1}^{(N)}$ being $N$ coproducts of the $U_{q}(sl_{2})$ generators, i.e.
\be K_{1}^{(N)} = \Delta^{(N)}(k_{1}), ~~~~E_{1}^{(N)} =\Delta^{(N)}(e_{1}),~~~~F_{1}^{(N)} =\Delta^{(N)}(f_{1}). \label{tp2} \ee We shall also need the asymptotic behavior of the ${\cal K}$ matrix which reads
\be {\cal K}(\lambda \to \infty)  \propto  \left( \begin{array}{cc}
e^{i\mu \xi} & \kappa  \\
 \kappa      &  
\end{array} \right).\label{k1} \ee 
Then gathering together the information provided by (\ref{T1}), (\ref{k1}) and bearing in mind the form of ${\cal T}$ we conclude that
\be {\cal T}(\lambda \to \infty) \propto {\cal T}^+ = 
 \left( \begin{array}{cc}
Q_{1}^{(N)}+x_{1}{\mathbb I} & (q-q^{-1})^{-1}{\mathbb I}  \\
 (q-q^{-1})^{-1}{\mathbb I}      &0 
\end{array} \right) \label{charge1} \ee where the explicit expression of the non-local charge $Q_{1}^{(N)}$ is given by 
\be Q_{1}^{(N)}= q^{-{1\over 2}}K_{1}^{(N)}E_{1}^{(N)}+q^{{1\over 2}}K_{1}^{(N)}F_{1}^{(N)}+x_{1} (K_{1}^{(N)})^{2}-x_{1} {\mathbb I}, \label{Q1} \ee with $x_{1} ={e^{i \mu \xi} \over 2\kappa \sinh i \mu}$. Notice that the first order of the above expansion gave rise to only one non-trivial generator, as a consequence the boundary quantum group possesses one generator, whose $N$ coproduct is provided by $Q_{1}^{(N)}$. It is clear that for $N=1$ one obtains the abstract generator of ${\mathbb B}(U_{q}(sl_{2}))$.

Let as elaborate a bit further on the boundary quantum group ${\mathbb B}(U_{q}(sl_{2}))$. As already mentioned,  ${\mathbb B}(U_{q}(sl_{2}))$ is defined by (\ref{re}) as $\lambda_{i} \to \infty$, (no $\lambda$ dependence, homogeneous gradation) i.e. \begin{equation} R^{\pm}_{12}\ {\cal T}_{1}^{+}\ \hat R_{12}^{+}\ {\cal T}_{2}^+ =  {\cal T}^{+}_{2}\  R_{12}^{+}\ {\cal T}_{1}^{+}\ \hat R_{12}^{\pm} \label{rr} 
\end{equation} where $\hat R^{\pm} ={\cal P} R^{\pm} {\cal P}$. In fact, the later formula (\ref{rr}) is equivalent to the quartic algebraic relation of the so called cylinder braid group, \be g_{0}\ g_{1}\ g_{0} \ g_{1} = g_{1}\ g_{0}\ g_{1}\ g_{0}. \label{braid2} \ee The blob algebra (\ref{TL}) is simply a quotient of the cylinder braid group, for a more detailed analysis we refer the reader to e.g. \cite{male, doma}. To obtain (\ref{rr}) we consider $g_{1} =h({\cal U}_{1})+q$, where $h({\cal U}_{1})$ is given in (\ref{tlg}), we act on (\ref{braid2}) with ${\cal P}$ from the left and right, and set $~R_{12}^{\pm} = {\cal P}_{12}\ g_{1}^{\pm 1}$, also ${\cal T}_{1}^{+}$ is a tensor representation of the generator $g_{0}$.  Relations (\ref{rr}) may be thought of as the boundary analogues of the well known relations for the upper lower Borel subalgebras defined by $R_{12}^{\pm} L_{1}^{+} L_{2}^{+} = L_{2}^{+}L_{1}^{+}R_{12}^{\pm}$... (see e.g. \cite{tak}). It is clear that for higher rank algebras there exist a set of boundary charges (non-affine) that form the boundary quantum group associated to the corresponding algebra. However, we repeat that in our case ${\mathbb B}(U_{q}(sl_{2}))$ is consisted by only one charge that is $Q_{1}$, which as will be shown in the last section is the centralizer of the blob algebra (\ref{TL}).\\
\\
$\bullet$ {\it Principal gradation}: The non-local charge derived in (\ref{Q1}) is associated to the non-affine generators of $U_{q}(\widehat{sl_{2}})$. To extract the charge associated to the affine generators it is convenient to consider the asymptotics of ${\cal T}$ in the principal gradation. Let us first note that ${\cal L}$ in the principal gradation may be obtained via the simple gauge transformation ${\cal V}$,  then as $\lambda \to \infty$ and bearing in mind the evaluation homomorphism (\ref{eval1}) we may express the ${\cal L}$ matrix in the following convenient form \be {\cal L}(\lambda \to \infty) \propto  \left(
\begin{array}{cc}
k_{1}  &    \\
&  k_{2}        \\
\end{array} \right ) +  e^{-\lambda} w  \left(
\begin{array}{cc}
&  f_{1}  \\
 f_{2}  &         \\
\end{array} \right), ~~~\hat {\cal L}(\lambda \to \infty) \propto  \left(
\begin{array}{cc}
k_{1}  &    \\
&  k_{2}        \\
\end{array} \right ) + e^{-\lambda} w\left(
\begin{array}{cc}
&  e_{2}  \\
 e_{1}  &         \\
\end{array} \right) \,.    \label{d0} \ee where $k_{2} =k_{1}^{-1}$. From (\ref{tt}), (\ref{th}) and (\ref{d0}) the asymptotics of $T$, $\hat T$ as $\lambda \to \infty$ can be easily obtained i.e.  \be && T(\lambda \to \infty) \propto \left(
\begin{array}{cc}
    K_{1}^{(N)}  &    \\
      &   K_{2}^{(N)} \\
\end{array} \right)+e^{-\lambda} w \left(
\begin{array}{cc}
 & F_{1}^{(N)}  \\
  F_{2}^{(N)}  &      \\
\end{array} \right) , \non\\ &&\hat T(\lambda \to \infty)  \propto \left(
\begin{array}{cc}
     K_{1}^{(N)}  &    \\
      &   K_{2}^{(N)} \\
\end{array} \right) +e^{-\lambda} w \left(
\begin{array}{cc}
    & E_{2}^{(N)}  \\
   E_{1}^{(N)}  &      \\
\end{array} \right).
\label{f2} \ee $K_{i}^{(N)}$, $E_{i}^{(N)}$ and $F_{i}^{(N)}$, provide $N$ coproduct of the quantum Kac--Moody algebra $U_{q}(\widehat{sl_{2}})$ (see also Appendix) \be K_{i}^{(N)} = \Delta^{(N)}(k_{i}), ~~~~E_{i}^{(N)} =\Delta^{(N)}(e_{i}),~~~~F_{i}^{(N)} =\Delta^{(N)}(f_{i}),~~~~i \in \{1,\ 2 \}. \ee
The ${\cal K}$ matrix as $\lambda \to \infty$ (\ref{k}) becomes (note that the constants $\xi$ and $\kappa$ appearing in (\ref{k}) are considered to be finite)
\be {\cal K}(\lambda \to \infty)  \propto \left(
\begin{array}{cc}
       &  \kappa  \\
       \kappa &    \\
\end{array} \right) +e^{-\lambda} \left(
\begin{array}{cc}
e^{i\mu \xi}   &   \\
               &- e^{-i\mu \xi}     \\
\end{array} \right) \,. \label{f0} \ee
Taking into account the asymptotics of $T$, $\hat T$ (\ref{f2}) and ${\cal K}$ (\ref{f0}), as $\lambda \to \infty$ and also (\ref{transfer}) one can derive the corresponding behaviour of ${\cal T}$ (\ref{transfer}), namely 
\be {\cal T}(\lambda \to \infty) \propto 
\left(
\begin{array}{cc}
       & 1 \\
    1  &     \\
\end{array} \right) +2 \sinh i\mu\ e^{-\lambda}\ \left(
\begin{array}{cc}
Q_{1}^{(N)}+x_{1}{\mathbb I} &  \\
                     &Q_{2}^{(N)}+x_{2} {\mathbb I}     \\
\end{array} \right) \label{asym} \ee where $x_{2} = -{e^{- i \mu \xi} \over 2\kappa \sinh i \mu}$ and the  non-local charges $Q_{i}^{(N)}$, $i\in\{1,\ 2 \}$ are given by the following expression \be Q_{i}^{(N)} = q^{-{1\over 2}}K_{i}^{(N)}E_{i}^{(N)}+q^{{1\over 2}}K_{i}^{(N)}F_{i}^{(N)}+x_{i} (K_{i}^{(N)})^{2}-x_{i} I, ~~~i\in \{ 1,\ 2\}. \label{Q} \ee The constant $x_{i}$ is subtracted in (\ref{Q}) so that convenient coproduct expressions can be obtained, as will become clear in the next section.  By keeping higher order terms one may attain higher conserved charges, this case however will be examined in detail elsewhere.

The boundary non--local charges entailed via algebraic considerations (see also \cite{dema}) are the same as the ones constructed  for the sine Gordon model on the half line \cite{mene, dema}. This is not a surprise since both boundary integrable field theories and discrete integrable models are ruled by the same set of algebraic constraints provided by (\ref{re}). This is the deep reason why the charges $Q_{1},\ Q_{2}$ agree with the ones obtained in \cite{mene, dema} in the context of the boundary sine--Gordon model. Our ultimate goal is to find conserved quantities commuting with the open transfer matrix, and this will be done in the subsequent sections.

\section{The symmetry}

In the previous section the boundary non--local charges were derived as coproducts of generators of the reflection algebra defined 
by (\ref{re}). It is clear that for $N=1$ we may obtain the corresponding abstract generators (see also \cite{mene, dema}) \be
{\cal Q}_{i} &=& q^{-{1\over 2}}k_{i}e_{i}+q^{ {1\over 2}}k_{i}f_{i} +x_{i} k_{i}^{2}-x_{i}{\mathbb I}, ~~~i \in \{1,\ 2\}. \label{genb} \ee  Let us denote $\hat {\mathbb B}=\{ {\cal Q}_{i} \} \subset {\mathbb R}$ \cite{dema, dege}. $\hat {\mathbb B}$ is endowed with a 
coproduct left from ${\cal A}$, with $\Delta: \hat {\mathbb B} \to \hat {\mathbb B} \otimes {\cal A}$. Taking into account the expressions (\ref{cop}), we may write the coproduct formed by ${\cal Q}_{i}$ in a convenient way as (see also \cite{dema, dege}) \be \Delta ({\cal
Q}_{i}) = {\mathbb I} \otimes  {\cal Q}_{i} + {\cal Q}_{i}\otimes \ k_{i}^{2}. \label{cop1} \ee Let us point out that the constant $x_{i}$ is subtracted in (\ref{genb}) so that a convenient coproduct form is attained. Without subtracting the constant we would have obtained a more complicated expression\footnote{ In fact, the corresponding coproduct is given by:\be \Delta({\cal Q}_{i}) = {\mathbb I} \otimes {\cal Q}_{i}+ {\cal Q}_{i} \otimes k_{i}^{2} - x_{i} {\mathbb I} \otimes k_{i}^{2} \ee}, nonetheless the results that follow are not affected by this choice in any way. 
We may now write the boundary non-local charges in a more compact form as \be Q_{i}^{(N)}=\Delta^{(N)}({\cal
Q}_{i}). \label{repb} \ee  It is worth remarking that recently it was shown \cite{base, base2} that the quantities (\ref{repb}) generate a so called tridiagonal algebra.

The remaining of this section is devoted to the derivation of intertwining relations involving the quantities (\ref{genb}) and the solutions of the reflection equation. Such relations are quite significant, because they enable the investigation of the exact symmetry of the open transfer matrix. Let $\rho_{\lambda}$ be the evaluation representation of $U_{q}(\widehat{sl_{2}})$ (see also Appendix) \cite{jimbo} $\rho_{\lambda}: U_{q}(\widehat{sl_{2}})\to \mbox{End}(\mathbb
C^{2})$ such that \be &&\rho_{\lambda}(k_{1}) =q^{ {\sigma^{z}\over 2}},~~~\rho_{\lambda}(e_{1})=
\sigma^{+},~~~\rho_{\lambda}(f_{1})=e^{
\lambda}\sigma^{-}, \non\\ &&\rho_{\lambda}(k_{2}) =q^{-{\sigma^{z}\over 2}},~~~\rho_{\lambda}(e_{2})=e^{-2
\lambda} \sigma^{-},~~~\rho_{\lambda}(f_{2})=e^{2
\lambda}\sigma^{+} . \label{action} \ee Notice that in this representation
$\rho_{\lambda}(k_{1})=\rho_{\lambda}(k_{2})^{-1} =q^{{1\over 2}\sigma^{z}}$, which implies that (\ref{action}) provides actually a deformation of the $\widehat{sl_{2}}$ loop algebra with zero center. From relations (\ref{genb}), (\ref{action}) it can be deduced that \be &&\rho_{\lambda}({\cal Q}_{1}) = q^{-{1\over 2}} q^{{ \sigma^{z}\over 2}} \sigma^{+}+ q^{{1\over 2}} q^{{\sigma^{z}\over 2}}
\sigma^{-}+ x_{1}q^{\sigma^{z}}-x_{1}{\mathbb I} \non\\&&\rho_{\lambda}({\cal Q}_{2}) = q^{-{1\over 2}}e^{-2
\lambda}q^{-{\sigma^{z}\over 2}} \sigma^{-}+ q^{{1\over 2}} e^{2\lambda} q^{-{\sigma^{z} \over 2}}
\sigma^{+}+ x_{2}q^{-\sigma^{z}}-x_{2}{\mathbb I}  . \label{Q2} \ee  It may be easily shown that all the elements of the reflection algebra  ${\mathbb K}_{ij}(\lambda)$ commute with the $c$-number solution of the reflection equation (see also \cite{dema}). 
Indeed, by acting with the evaluation representation on the second space of 
(\ref{gensol}) we obtain \be (\mbox{id} \otimes \rho_{\pm \lambda}) {\mathbb K}(\lambda') = R(\lambda' \mp \lambda)\ ({\cal K}(\lambda') \otimes {\mathbb I})\ \hat
R(\lambda' \pm\lambda), \ee where $\hat R(\lambda) ={\cal P}\ R(\lambda)\ {\cal P}$. Now recalling 
the reflection equation (\ref{re}) and because of the form of the  above expressions it is straightforward to show that the entries of  ${\mathbb K}$ in the evaluation representation `commute' with the c-number ${\cal K}$ matrix (\ref{k}) \be 
\rho_{\lambda}({\mathbb K}_{ij}(\lambda'))\ {\cal K}(\lambda) = {\cal K}(\lambda)\ \rho_{-\lambda}({\mathbb K}_{ij}(\lambda')), ~~~i,\ j \in \{1, \ 2 \}. \label{bcomm} \ee  Recall however that ${\mathbb K}_{ij}(\lambda \to \infty)$ provide essentially the generators (\ref{genb}), and thus it immediately follows that
\be \rho_{\lambda}({\cal Q}_{i})\ {\cal K}(\lambda)={\cal K}(\lambda)\ \rho_{-\lambda}({\cal Q}_{i}).
\label{ik} \ee The above relations have been also verified by inspection considering the explicit form of (\ref{Q2}), (\ref{k}).

The latter equations (\ref{ik}) are the boundary analogues of the well known `commutation' relations for the Lax operators \be &&(\rho_{\lambda} \otimes \mbox{id} )\Delta'(x)\ {\cal L}(\lambda) = {\cal L}(\lambda)\ (\rho_{\lambda} \otimes  \mbox{id} ) \Delta (x)\, \non\\ &&(\rho_{-\lambda} \otimes \mbox{id} )\Delta(x)\ \hat {\cal L}(\lambda) =  \hat {\cal L}(\lambda)\ ( \rho_{-\lambda} \otimes  \mbox{id} ) \Delta'(x) ~~~~~x\in \{e_{i},\ f_{i},\ k_{i} \}.\label{ir} \ee Relations of the type (\ref{ir}) were derived
originally in \cite{kure, jimbo2}, and also used extensively for deriving $R$ matrices and exact $S$ matrices in 2D quantum integrable models \cite{kure, bele}. On the other hand relations (\ref{ik}) were employed in e.g. \cite{dema}--\cite{bako} for obtaining solutions of the reflection equation (\ref{re}) associated to a variety of integrable models.

The main objective now is the derivation of generalized intertwining relations of the type (\ref{ik}), (\ref{ir}) for ${\cal T}$. Indeed it turns out via (\ref{ik}), (\ref{ir}) that the following generalized intertwining relations hold \be (\rho_{\lambda} \otimes \mbox{id}^{\otimes N})\Delta^{'(N+1)}({\cal Q}_{i})\ {\cal T}(\lambda) = {\cal T}(\lambda)\ (\rho_{-\lambda} \otimes \mbox{id}^{\otimes N})\Delta^{'(N+1)}({\cal Q}_{i}). \label{it} \ee Equations (\ref{it}) are of great relevance bearing explicit algebraic relations among the entries of the ${\cal T}$ matrix and the non-local charges (\ref{Q}) (see \cite{kor} for the bulk analogue), allowing the derivation of the symmetry of the open spin chain. We shall derive the symmetry of the transfer matrix in the homogeneous gradation, recall however that the two gradations are related via a simple gauge transformation (\ref{gauge1}). Let \be {\cal T}(\lambda) = \left(
\begin{array}{cc}
{\cal A}_{1} &  {\cal B}\\
{\cal C} &  {\cal A}_{2}   \\
\end{array} \right)  \label{tr2}
\ee then using (\ref{cop1}), (\ref{Q2}), (\ref{it}), (\ref{tr2}) and defining $\eta_{i} =(-1)^{i+1}$ the following important algebraic relations are entailed \be &&\Big [Q_{i}^{(N)},\ {\cal A}_{1} \Big ]= 
e^{\lambda-\eta_i (\lambda +i\mu)} ({\cal B}-{\cal C}),~~~~\Big [ Q_{i}^{(N)},\ {\cal A}_{2} \Big ]= -e^{-\lambda+\eta_{i}(\lambda +i\mu)}({\cal B}-{\cal C}) \non\\ 
&& \Big [ Q_{i}^{(N)},\ {\cal C} \Big ]_{q^{-\eta_{i}}} = 
e^{\lambda-\eta_{i}\lambda}
{\cal A}_{2}- e^{-\lambda+\eta_{i} \lambda} {\cal A}_{1} +x_{i}(q^{\eta_{i}}- q^{-\eta_{i}}){\cal C} \non\\
&& \Big [Q_{i}^{(N)},\ {\cal B} \Big ]_{q^{\eta_{i}}} =
e^{-\lambda+\eta_{i}\lambda} {\cal A}_{1}- e^{\lambda-\eta_{i}\lambda} {\cal A}_{2} +x_{i}(q^{-\eta_{i}}- q^{\eta_{i}}){\cal B}, ~~~i\in \{1,\ 2 \} \label{com2} \ee where we define $[X,\ Y]_{q} =qX Y-q^{-1}Y X$. The algebraic relations (\ref{com2}) together with (\ref{it}) are of the most important results of this article. 
Equations (\ref{com2}) are essentially algebraic Bethe ansatz type commutation relations similar to the ones appearing in \cite{sklyanin}. We shall now consider three different choices for ${\cal K}^{(l)}$:\\
\\
{\bf (I)}  ${\cal K}^{(l)}(\lambda) ={\mathbb I}$: Recall that in the homogeneous gradation $M =diag(q,\ q^{-1})$, and the transfer matrix can be now written as: \be t(\lambda) = e^{ i \mu} {\cal A}_{1} + e^{- i\mu} {\cal A}_{2}. \label{tt2} \ee Then by virtue of (\ref{tt2}) and recalling the exchange relations (\ref{com2}) between the boundary non-local charges and the entries ${\cal A}_{i}$ it follows that:
\be \Big [t(\lambda),\ Q_{1}^{(N)} \Big ]= 0, ~~\Big [t(\lambda),\ Q_{2}^{(N)} \Big ]=-2 \sinh 2(\lambda+i\mu)  ({\cal B}- {\cal C}). \label{co2} \ee
\\
{\bf (II)}  ${\cal K}^{(l)}(\lambda) = diag(e^{-2\lambda -2 i\mu},\ e^{2\lambda +2 i\mu})$: Note that this solution follows from (\ref{k}), in particular ${\cal K}^{(l)}(\lambda)= {\cal K}(-\lambda -i\mu;\ i\mu\xi \to \infty)$. Similarly, by means of (\ref{hh}) the transfer matrix in the homogeneous gradation can be now written as: 
\be  t(\lambda) = e^{-2\lambda - i \mu} {\cal A}_{1} + e^{2\lambda + i\mu} {\cal A}_{2} \label{tt3} \ee and via (\ref{com2}) and (\ref{tt3}) we conclude:
\be \Big [t(\lambda),\ Q_{1}^{(N)} \Big ]= 2 \sinh 2(\lambda+i\mu)  ({\cal B}- {\cal C}), ~~\Big [t(\lambda),\ Q_{2}^{(N)} \Big ]=0. \label{co3} \ee 
\\
{\bf (III)}  Let finally ${\cal K}^{(l)}(\lambda) = diag(e^{-\lambda -i\mu},\ e^{\lambda + i\mu})$. Then the transfer matrix may be written as \be t(\lambda) = e^{-\lambda } {\cal A}_{1} + e^{\lambda} {\cal A}_{2} \label{tt4}  \ee From (\ref{tt4}), (\ref{com2}) it immediately follows \be \Big
[t(\lambda),\ Q_{1}^{(N)}\Big ] = - 2\sinh(\lambda+i\mu) ({\cal B}- {\cal C}), ~~\Big [ t(\lambda),\ Q_{2}^{(N)}\Big ] = 2\sinh(\lambda+i\mu) ({\cal B}- {\cal C}) \ee and we conclude that \be \Big [t(\lambda),\ Q_{1}^{(N)}+Q_{2}^{(N)} \Big ] =0. \label{co1} \ee
We arrived at these results by simply exploiting the generalized intertwining relation (\ref{it}).  The procedure described in the present section provides a convenient way for studying the symmetry of open spin chains, and this is the first time to our knowledge that this method is used in this context.

Relations similar to (\ref{com2}) may be deduced for the case where ${\cal T}$ is in the principal gradation.  The transfer matrix in the principal gradation is (\ref{transfer}) \be t^{(p)}(\lambda) = Tr_{0}\ \Big \{ M_{0}\ {\cal T}_{0}^{(p)}(\lambda) \Big \} \label{hh} \ee where ${\cal T}^{(p)}$ is given by (\ref{th}) and recall that in the principal gradation $M ={\mathbb I}$. To be more specific let us write down the expressions for the left boundary of the transfer matrix in the principal gradation for the cases described in (I), (II), (III) obtained trivially via the gauge transformation (\ref{gauge2}): \be && \mbox{(I)} ~~~~{\cal K}^{(l,\ p)}(\lambda) = diag(e^{\lambda +i\mu},\ e^{-\lambda - i\mu}) \non\\  && \mbox{(II)} ~~~~{\cal K}^{(l,\ p)}(\lambda) = diag(e^{-\lambda -i\mu},\ e^{\lambda + i\mu}) \non\\  && \mbox{(III)} ~~~~{\cal K}^{(l,\ p)}(\lambda) ={\mathbb I}. \ee The right boundary in the principal gradation is given by \be {\cal K}^{(r,\ p)}(\lambda) = \left(
\begin{array}{cc}
\sinh(\lambda +i\mu \xi)   & \kappa\ \sinh 2 \lambda  \\
\kappa\ \sinh 2\lambda  &\sinh(-\lambda +i\mu \xi)     
\end{array} \right) \label{kpr}. \ee Note that the transfer matrix of the system remains invariant under the gauge transformation (\ref{gauge1}), what changes only is the form of the $R$ and ${\cal K}$ matrices from one gradation to the other.

A more general diagonal (see e.g. \cite{pasa}--\cite{done}) or non-diagonal left boundary could have been applied (see also \cite{base2}). In the case of a general diagonal left boundary the transfer matrix would be expressed as linear combination of the entries ${\cal A}_{i}$, whereas for a non-diagonal boundary the transfer matrix would be written as combination of all the entries of ${\cal T}$ i.e. ${\cal A}_{i}$, ${\cal B}$ and ${\cal C}$. In both cases we are able to derive the commutators between the boundary non-local charges and the transfer matrix due to relations (\ref{com2}). Furthermore, as was shown in \cite{base2} $Q_{1}^{(N)}$ and $Q_{2}^{(N)}$ give rise to a hierarchy of charges commuting  with the transfer matrix, and consisting an abelian algebra. All the higher charges may be written as combinations of the fundamental generators $Q_{1}^{(N)}$ and $Q_{2}^{(N)}$. As a consequence, knowing the exchange relations (\ref{com2}) we may check all the possible combinations of these two charges commuting with the transfer matrix for various choices of boundary conditions. This is the first time that relations of the type (\ref{com2}), and hence the symmetry (\ref{co2}), (\ref{co3}), (\ref{co1}), are deduced for an integrable open spin chain with non-diagonal boundary conditions, a fact that clearly indicates the importance of these findings. The generalized intertwining relations (\ref{it}) as well as the exchange relations (\ref{com2}), and all the discovered symmetries (\ref{co2}), (\ref{co3}), (\ref{co1}) are universal results since they have been derived independently of the choice of representation.
  
\subsection*{Commutation with the blob algebra} 

In what follows a connection between the boundary quantum group generator (\ref{Q}) and the generators of the blob algebra will be made. More specifically, we shall show that the representation of the non-local charge $\rho_{0}^{\otimes N}(Q_{1}^{(N)})$ (\ref{Q}) commutes with each one of the blob algebra generators  \cite{mawo, doma} in the XXZ representation (\ref{tlg}), i.e.  \be \Big [h({\cal U}_{0}),\ \rho_{0}^{\otimes N}(Q_{1}^{(N)}) \Big ]=0. \label{a} \ee It can be shown  by inspection by virtue of (\ref{tp2}), (\ref{Q}) (for $N=1$), (\ref{action}), (\ref{tlg}), (\ref{ide}) that (\ref{a}) is valid for $N=1$. Then by means of (\ref{cop1}), (\ref{repb}), and  (\ref{tlg}) it follows that (\ref{a}) is valid for any $N$. Indeed, notice from (\ref{cop1}) (for $i=1$) that the first site of the coproduct is occupied either by the unit element or by ${\cal Q}_{1}$, representations of which on ${\mathbb C}^{2}$ commute with $h({\cal U}_{0})$ (\ref{tlg}) as checked by inspection for $N=1$.

It is also well known (see e.g \cite{pasa}) that the generators of the Temperley-Lieb algebra in the XXZ representation $h({\cal U}_{l})$ (\ref{tlg})
commute with the quantum group generators (\ref{tp2}), i.e. \be \Big [h({\cal U}_{i}),\ \rho_{0}^{\otimes N}(E_{1}^{(N)}) \Big ]=\Big [h({\cal U}_{l}),\ \rho_{0}^{\otimes N}(F_{1}^{(N)}) \Big ]=\Big [h({\cal U}_{l}),\ \rho_{0}^{\otimes N}(K_{1}^{(N)}) \Big ]=0, ~~l\in \{1,...,N-1 \}
\label{b} \ee and by virtue of (\ref{Q}) it is obvious that \be \Big [h({\cal U}_{l}),\ \rho_{0}^{\otimes N}(Q_{1}^{(N)}) \Big ]=0. \label{c} \ee From equations (\ref{a}), (\ref{c}) it is immediately entailed that all the generators of the blob algebra, in the XXZ representation, commute with the non-local charge $Q_{1}^{(N)}$ (\ref{Q}), \be \Big [h({\cal U}_{l}),\ \rho_{0}^{\otimes N}(Q_{1}^{(N)}) \Big ]=0, ~~l\in \{0,1,...,N-1 \}.\label{e} \ee The commutation relations (\ref{e}), which are also among the main results of this work, are the boundary analogues of (\ref{b}) and they can be used for studying the symmetry of the open XXZ spin chain Hamiltonian as e.g. in \cite{pasa}. 

In particular, let us focus on the special case where the fundamental representation (\ref{action}) acts on the quantum spaces of the spin chain (${\cal L} \to R$), and consider ${\cal K}^{(l)}(\lambda) = {\mathbb I}$, $~{\cal K}^{(r)}(\lambda)$ is given by (\ref{ansatz1}) (homogeneous gradation).
The corresponding open spin chain Hamiltonian may be then expressed as \cite{done}, \be {\cal H} = -{(\sinh i\mu)^{-2N+1} \over {4 x(0)}} \Big( tr_{0} M_{0} \Big )^{-1}\ 
\ {d \over d \lambda} t(\lambda)\vert_{\lambda =0} \label{H0}. \ee  Recall also that $R$ is given by (\ref{ansatz2}). Then having in mind that  $~~R(0)= \hat R(0)=\sinh i \mu\ {\cal P}, ~~~{\cal K}^{(r})(0) = x(0)\ {\mathbb I}~~$, and after some algebra we conclude that the Hamiltonian (\ref{H0}) may be written as
\be  {\cal H} = -\frac{1}{2}
\sum_{l=1}^{N-1} h({\cal U}_{l}) - \frac{\sinh i\mu\ y'(0)}{4  x(0)} h({\cal U}_{0})+ w 
\label{ht} \ee where $w = -{\sinh i \mu x'(0) \over 4 x(0)}-{N\over 2}\ \cosh i \mu + {1\over 2(q+q^{-1})}$. The latter Hamiltonian reduces to (\ref{Hbound}) after expressing the blob generators in terms of the Pauli matrices, and for this particular choice of the left boundary the constants in (\ref{Hbound}) become $c_{1}={1\over 2(q+q^{-1})}, \ c_{2} =0$.  Note that for the values $\zeta = -{m\over 2}$, $~\zeta = {m\over 2}+{\pi \over 2 \mu}\  ~(mod({\pi \over \mu}))$, already mentioned in the introduction, $y'(0) =0$, and thus there is no contribution from the blob generator ${\cal U}_{0}$.

The Hamiltonian (\ref{ht}) is solely expressed in terms of the blob algebra generators in the XXZ representation (\ref{tlg}), and as a consequence of (\ref{e}) we conclude \be \Big [{\cal H}, \ \rho_{0}^{\otimes N}(Q_{1}^{(N)}) \Big ]=0. \label{ham} \ee The latter commutation relation could have been obtained by means of (\ref{co2}) and via (\ref{H0}), by simply acting on the quantum spaces with the evaluation representation. Of course the crucial point here is not only that the open Hamiltonian (\ref{ht}) commutes with $\rho_{0}^{\otimes N}(Q_{1}^{(N)})$, but more importantly that each one of the generators of the blob algebra commute with it. It should be stressed that in the present study we deal with a somehow trivial left boundary (see cases (I), (II) and (III)). In this case an interesting analysis on the spectrum of the system, as well as on the spectrum of the aforementioned non-local charge $Q_{1}^{(N)}$, for various values of the parameters of the blob algebra, is presented in \cite{nic}. If on the other hand one considers a non-trivial left boundary, then an extra generator needs to be added to the blob algebra and one has to proceed as in \cite{degi, degi2}. This case although of great interest is not pursued in the present work.
   
\section{Discussion}

Given a ${\cal K}$ matrix, solution of the reflection equation (\ref{re}), one can explicitly construct the boundary non-local charges (\ref{Q}) via the asymptotic behaviour of ${\cal T}$ as $\lambda \to \infty$ (see also \cite{dema}). Note that solutions of the reflection equation (\ref{re}) may be available by means of a variety of techniques such as e.g. the Hecke algebraic approach \cite{doma}. One then may identify linear intertwining relations between them and the solutions of the reflection equation (\ref{ik}), (\ref{it}), which turn out to be of great significance, because they provide algebraic relations (\ref{com2}) between the entries of the transfer matrix and the non-local charges (\ref{Q}), facilitating the study of the transfer matrix symmetry with special left boundary (\ref{co1}), (\ref{co2}), (\ref{co3}). In addition, we were able to show that one of the non-local charges (\ref{Q}), in a certain representation, commutes with the XXZ realization of the blob algebra generators for a particular choice of the left boundary. This fact enables the investigation of the symmetry of the corresponding local Hamiltonian (\ref{ham}), written exclusively in terms of the blob algebra generators.

The procedure applied here may be generalized in a straightforward manner for models associated to higher rank algebras such as e.g $A_{n-1}^{(1)}$, (see also \cite{neg}). In particular, after extracting the boundary non-local charges via the standard process the main point is to obtain generalized relations of the type (\ref{com2}), study the symmetry of the corresponding open transfer matrix, and also prove the existence of centralizers of the corresponding affine Hecke algebra.\\
\\
\textbf{Acknowledgements:} This work is supported by the TMR Network `EUCLID.  Integrable models and applications: from strings to condensed matter', contract number HPRN-CT-2002-00325.

\appendix

\section{The quantum Kac--Moody algebra $U_{q}(\widehat{sl_{2}})$}

It is instructive to briefly review some basic definitions concerning the quantum group structures. Let \be (a_{ij}) =
\left(
\begin{array}{cc}
   2   &-2\\
  -2  &2\\
\end{array} \right)
\,\ee be the Cartan matrix of the affine Lie algebra
${\widehat{sl_{2}}}$ \cite{kac}, and also define \be [x]_{q} = {q^{x} -q^{-x} \over q-q^{-1}}. \ee Recall that the quantum affine enveloping algebra $U_{q}(\widehat{sl_{2}})\equiv \cal A$ has the Chevalley-Serre generators \cite{jimbo, drinf}  $e_{i}$, $f_{i}$, $k_{i}$, $i\in \{1,\ 2\}$ obeying the defining relations  \be k_{i}\ k_{j} = k_{j}\ k_{i}, ~~k_{i}\ e_{j}&=&q^{{1\over
2}a_{ij}}e_{j}\ k_{i}, ~~k_{i}\ f_{j}
= q^{-{1\over 2}a_{ij}}f_{j}\ k_{i}, \non\\
\Big [e_{i},\ f_{j}\Big ] &=& \delta_{ij}{k_{i}^{2}-k_{i}^{-2} \over q-q^{-1}}, ~~i,j \in \{ 1,\ 2 \} \label{1}\ee and the $q$ deformed Serre relations
\be \chi_{i}^{3}\ \chi_{j} -[3]_{q}\ \chi_{i}^{2}\ \chi_{j}\ \chi_{i} +[3]_{q}\ \chi_{i}\ \chi_{j}\ \chi_{i}^{2} -\chi_{j}\ \chi_{i}^{3} =  0, ~~~~\chi_{i} \in \{e_{i},\ f_{i} \}, ~~~i\neq j. \ee There exists a homomorphism called the evaluation homomorphism \cite{jimbo} $\pi_{\lambda}: U_{q}(\widehat{sl_{2}}) \to U_{q}(sl_{2})$ \be &&\pi_{\lambda}(e_{1}) = e_{1}, ~~~~~\pi_{\lambda}(f_{1}) =f_{1}, ~~~~\pi_{\lambda}(k_{1}) = k_{1} \non\\ && \pi_{\lambda}(e_{2}) = e^{-2\lambda}c f_{1}, ~~~~~\pi_{\lambda}(f_{2}) =e^{2\lambda}c^{-1}e_{1}, ~~~~\pi_{\lambda}(k_{2}) = k^{-1}_{1}.\label{eval1} \ee where $c$ is a constant.
As mentioned this algebra is also equipped with a coproduct $\Delta: {\cal A}
\to {\cal A} \otimes {\cal A}$, in particular the generators form the following coproducts \be &&\Delta(y) = k_{i}^{-1} \otimes y + y \otimes k_{i}, ~~~y\in \{e_{i},\ f_{i} \} ~~~\mbox{and} ~~~\Delta(k_{i}^{\pm
1}) = k_{i}^{\pm 1} \otimes k_{i}^{\pm 1}. \label{cop} \ee

\end{document}